\documentclass[runningheads]{llncs}
\usepackage{graphicx}
\usepackage{listings}
\usepackage{color}
\usepackage{comment}
 
\definecolor{codegreen}{rgb}{0,0.6,0}
\definecolor{codegray}{rgb}{0.5,0.5,0.5}
\definecolor{codepurple}{rgb}{0.58,0,0.82}
\definecolor{backcolour}{rgb}{1.0,1.0,1.0}
\lstdefinestyle{mystyle}{
    backgroundcolor=\color{backcolour},   
    commentstyle=\color{codegreen},
    keywordstyle=\color{magenta},
    numberstyle=\tiny\color{codegray},
    stringstyle=\color{codepurple},
    basicstyle=\footnotesize\ttfamily,
    breakatwhitespace=false,         
    breaklines=true,                 
    captionpos=b,                    
    keepspaces=true,                 
    numbers=left,                    
    numbersep=5pt,                  
    showspaces=false,                
    showstringspaces=false,
    showtabs=false,                  
    tabsize=2
}
\usepackage{array}
\newcommand{\PreserveBackslash}[1]{\let\temp=\\#1\let\\=\temp}
\newcolumntype{C}[1]{>{\PreserveBackslash\centering}p{#1}}
\newcolumntype{R}[1]{>{\PreserveBackslash\raggedleft}p{#1}}
\newcolumntype{L}[1]{>{\PreserveBackslash\raggedright}p{#1}}

\usepackage{arydshln}
\usepackage{cite}
\usepackage{todonotes}

\begin{document}
\title{Modernizing Titan2D, a Parallel AMR Geophysical Flow  Code to Support Multiple Rheologies and Extendability\thanks{We gratefully acknowledge the support of NSF awards OAC 1339765}}
%
%
\author{Nikolay~A. Simakov\inst{1}\orcidID{0000-0001-6131-5979} \and
Renette~L. Jones-Ivey\inst{1,2}\orcidID{0000-0002-0907-7624} \and
Ali Akhavan-Safaei\inst{2}\orcidID{0000-0002-0812-1881} \and
Hossein Aghakhani\inst{2}\orcidID{0000-0003-2398-9301} \and
Matthew~D. Jones\inst{1}\orcidID{0000-0001-7293-226X} \and
Abani~K. Patra\inst{1,2}\orcidID{0000-0002-3571-5212}} 
\authorrunning{N.A. Simakov et al.}
\titlerunning{Modernizing Titan2D}
%
\institute{
Center for Computational Research, 
 University at Buffalo, Buffalo,NY\and
Computational and Data-Enabled Sciences and Engineering, 
 University at Buffalo, Buffalo, NY\\
\email{abani@buffalo.edu}
}

\maketitle              
\begin{abstract}
In this work, we report on strategies and results of our initial approach for modernization of Titan2D code. Titan2D is a geophysical mass flow simulation code designed for modeling of volcanic flows, debris avalanches and landslides over a realistic terrain model. It solves an underlying hyperbolic system of partial differential equations using parallel adaptive mesh Godunov scheme. The following work was done during code refactoring and modernization. To facilitate user input two level python interface was developed. Such design permits large changes in C++ and Python low-level while maintaining stable high-level interface exposed to the end user. Multiple diverged forks implementing different material models were merged back together. Data storage layout was changed from a linked list of structures to a structure of arrays representation for better memory access and in preparation for further work on better utilization of vectorized instruction. Existing MPI parallelization was augmented with OpenMP parallelization. The performance of a hash table used to store mesh elements and nodes references was improved by switching from a linked list for overflow entries to dynamic arrays allowing the implementation of the binary search algorithm. 
The introduction of the new data layout made possible to reduce the number of hash table look-ups by replacing them with direct use of indexes from the storage class. The modifications lead to 8-9 times performance improvement for serial execution.

\end{abstract}

\section{Introduction}

Titan2D is a tool developed for simulation of granular flows over digital elevation models of natural terrain including volcanic flows, landslides, debris and snow 
avalanches \cite{Pitman2003,Patra2005,patra_nishimura2018}. It is widely used by researchers and civil protection authorities for risk assessment and mitigation scenarios for these hazards. 

The Titan2D code built on an earlier AMR infrastructure developed for fluid dynamics simulations using adaptive $hp$ finite elements \cite{Laszloffy:2000:SDM:367574.367627}. While this allowed rapid and reliable development, it also baked in design decisions in data structures and coding that were sub-optimal. Secondly, the wide adoption of the code based on its ability to use sophisticated computing methodologies like AMR and its application to many different physical contexts led to many versions and forks.

Many choices in the software design are motivated by how the computational problem is mapped to available hardware, therefore it is not surprising that many legacy applications do not operate optimally on modern systems. In the case of Titan2D many currently sub-optimal designs decisions were largely motivated by a file system input/output bottleneck caused by the inability to hold the whole topography in the available system memory. Since the I/O performance limitation disguised other inefficiencies, many design decisions favored ease of development or debugging. For example,  a hash values lookup table (hash function based on space filling curve (SFC) ordering of cells) was used for accessing neighboring elements every time paying extra penalty for a search. However, it simplified the development as there is no need to track memory location of neighboring elements for each element and the performance penalty was way below that caused by poor filesystem performance. Nowadays most of the topological maps can fit into main memory and the element search penalty significantly degrades the overall performance.

Titan2D is open source and distributed through GitHub as source code and binaries. Code consists of over 50 thousand lines (generated using David A. Wheeler's SLOCCount) and is primarily written in C++, with new interface in Python and GUI written in Java. Titan2D is also available through an online resource for collaboration in volcanology research and risk mitigation called Vhub (vhub.org). It is a turnkey solution for running Titan2d. Since 2011 there were 650 unique users from 8 different counties (about 100 active users each year) executing more than 5 thousand simulation runs. Because GitHub does not provide enough statistic on utilization (it tracks cloning and visits for only two weeks), we cannot estimate the number of users who obtained code through GitHub and running it on their system. In our HPC center, all jobs executed Titan2D consume more than 2 million CPU hours (257 CPU years) since 2015 (tracking with XDMoD). Thus, performance and usability improvement benefit not only the developers' team and close affiliates but also other users as well.

In this work we report on our initial approach to modernize Titan2D. We replace the fixed format input with a Python interface, redesigned the data structure layout to benefit vectorized instructions, combined multiple forks under the same codebase, reduced hash table use with direct index use and introduced hybrid OpenMP/MPI parallelization. In the next section we briefly describe the Titan2D governing equation and numerical methods it uses to solve them. In the following sections we described our modifications in more detail.

\begin{figure}[t]
\includegraphics[width=\textwidth]{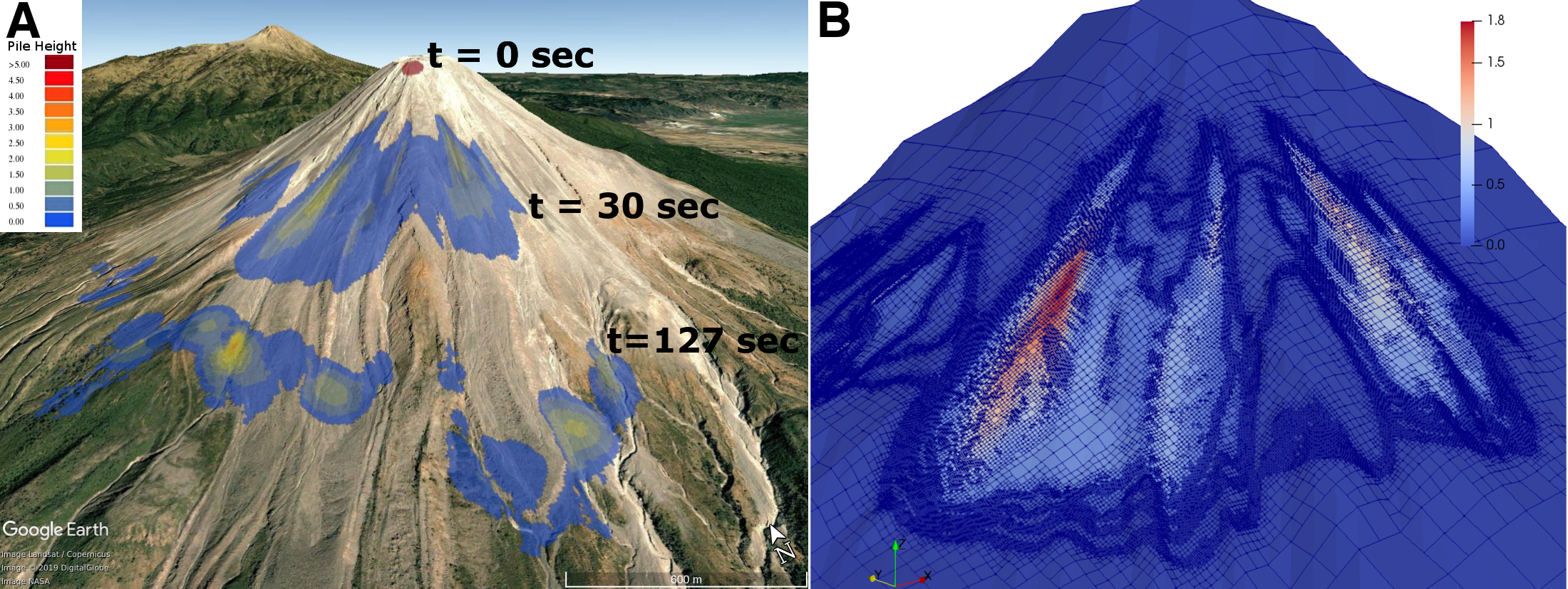}
\caption{Simulation of block and ash flows that have occurred at the Colima volcano, Mexico, during the 1991 year eruptive episode. \textbf{A}. Overlay of three snapshots from the simulation visualized with Google Earth. \textbf{B}. Illustration of adaptive grid used for same simulation, the snapshot corresponds to the middle region from \textit{A}, visualized with ParaView. } \label{figColima}
\end{figure}

\section{Titan2D and Benchmark Problem}

Titan2D is based upon a depth-averaged model for an incompressible continuum, a “shallow-water” granular flow. We will briefly review the functionality here; details of the modeling and numerical methodology may be found in the literature~\cite{Pitman2003,Patra2005}. The conservation equations for mass and momentum are solved with different rheologies modeling the interactions between the grains of the media and between the granular material and the basal surface. The resulting hyperbolic system of equations is solved using an MPI parallel, adaptive mesh (see Fig.~\ref{figColima}.B for adaptive grid illustration), Godunov scheme. Titan2D also takes as input the topography maps in the form of Digital Elevation Models (DEM) and (where available) local information on basal resistance to the flow.
Source codes, examples and Linux binaries are available at the Titan2D GitHub repository (https://github.com/TITAN2D).

\textbf{Benchmark Problem.} To measure the overall performance impact from our modifications, block and ash flow simulations on the Colima volcano, Mexico, (see Fig.~\ref{figColima}) were done for the legacy and modernized versions. The simulations were done with Coulomb material model, first order method with AMR. The calculation was done for 10000 iterations totalling 2 minutes total flow time. The average number of elements was 141 thousand. This simulation corresponds to one from the Titan2D examples with increased space accuracy (number of cells across smallest pile axis increased from 16 to 32). This problem is of lower middle size. Number of users have done much larger computations while larger portion of users perform calculations on similar or smaller sizes. Thus, for the purpose of analysis this problem size exercises the code well.  The legacy code was executed in MPI mode while modernized in OpenMP mode. 
Calculations were done on two systems: 1) desktop  (Intel Core i7-9800X Skylake-X CPU, 8 cores, 3.80 GHz; 64 GB of RAM (4xDDR4-2800); 1TB SATA SSD) and 2) academic compute cluster at the Center for Computational Research, University at Buffalo, SUNY (Intel Xeon Gold 6130, Skylake-X CPU,2x16 cores, 2.10GHz;	192GB of RAM; Isilon file-system).

\section{Refactoring Strategies}
\subsection{Adopting a Python Interface}

Legacy Titan2D has a fixed input file format, where each line specifies a particular parameter. Such an input system is easy to implement, however it has many drawbacks and lacks flexibility and is hard to extend to new models and options. Unused  parameters still should be specified even when unused making software hard to maintain. Nevertheless, ease of implementation made this choice popular in older software especially given harder to find and steeper learning curves to other C/C++ implemented alternatives. In our modernization effort the fixed input format was replaced with a Python interface.

A Python interface to C++ applications provides many benefits. It is highly flexible and allows the implementation of multiple methods and models under the same codebase as well as a number of ways to extend the application. Its programmability allows the creation of higher-level simulations schemes like parameter search and better integration with a Graphical User Interface (GUI). It also has a number of ways to interface with programs and libraries written in C, C++ and Fortran. Due to these benefits Python is used in many scientific applications and chosen for modernization or to serve as "glue" for multiple legacy applications~\cite{Marais2007,ladwig2017wrf,Krischer2015}. All this makes Python very popular and many modelers are familiar with it.
It also has a large ecosystem of scientific libraries and applications. Thus, the implementation of the Titan2D Python interface would allow closer integration with that ecosystem and reuse of already developed software.  

There are a number of ways for creating a Python API for C++ application. We choose to go with the Simplified Wrapper and Interface Generator (SWIG) since it provides automatic mapping of C++ classes and functions to Python. Such a mapping results in a Python API which closely follows C++ API. Due to the anticipation that the C++ API would change drastically during modernization, we chose two level design for Python interface: the low-level Python interface is automatically generated by SWIG and high-level is built on top of it and is exposed to the end user. This allows us to stabilize the user exposed high-level interface earlier while giving us the freedom of unrestricted modifications of underlying C++ and low-level Python interface.

Python programming language combined with C++ allows us to use the strengths of both languages. Python has significantly less boilerplate allowing fast prototyping and development while C++ can achieve  high performance. Thus, using Python in performance non-critical places would lead to faster development. This way we also significantly improve user interaction by implementing extensive user input validation in the high-level Python layer.

\subsection{Merging Multiple Forks}

The rigidity of the fixed input format resulted in multiple Titan2D forks which differ in physical model or numeric method. Pressed by deadlines or the desire to test promising new models led some researchers to replace parts of code with a new implementation instead of extending the original code. This lead to multiple forks which largely differ only in few places where the new model replaced the original one. Because forks were branched at different times the forked version also differs in some core Titan2D classes and functions which were changed after the fork. Maintaining multiple forks implementing different methods in multiple places is a labor-intensive task even using modern tools with sophisticated branch merging routines. Another drawback of multiple forks is that improvements to core functionality in one fork do not automatically  transfer to other forks. The Python interface provides the flexibility to keep multiple models under the same code base. In new Titan2D, the model and its parameter are specified by user using high-level Python interface and the parameters are propagated down to the C++ level. Because of forking at different times, code replacement for new models and lack of multi-model support  in the original code an automatic merge was not possible. Consequently, the merging was done manually with heavy utilization of Integrated Developer Environment (IDE) code analysis tools and stand-alone difference visualization and merging tools.

The first merged fork was one implementing two-phase Pitman-Le material model~\cite{Pitman2005} with the original single phase model. Due to the need to describe two separate phases this fork has significant differences with original Titan2D and the following strategy was used. First two branches were merged together into a single source code base and functionally similar parts were placed next to each within a conditional compilation if-else macro. That is, if the COULOMB\_TWO\_PHASE macro was defined the resulting code produced two phase Pitman-Le material model and the original single phase Coulomb model otherwise. Such single code base allows to see clearly all changes associated with merging fork and identify the best strategy to support both models without recompilation. To aid  merging the visual differences and merge tool called Meld~\cite{Willadsen_MeldTool} was used. 
Next, the parameters needed for new models were identified and new model selection and parameters specification were implemented in Python with their propagation to the C++ level. Finally, the conditional compilation if-else macro was replaced either with conditions or separate functions depending on the amount and type of differences.
Two other forks implementing Voellmy-Salm~\cite{Christen2010} and Pouliquen-Forterre~\cite{Pouliquen2002, Forterre2008} rheologies had very small and isolated changes and did not required as extensive work as the previously described merge. Briefly, the Python interface was extended to enable setting new parameters and propagate them to C++; next, several functions was converted from Fortran to C++, re-named and copied to master Titan2D.
The converted Fortran function performed a computation for a single cell; thus it did not provide benefits of Fortran compiler optimization and converting to C++ inline function also allows to eliminate functional call and add space for further optimization.

The lesson learned: do not allow forks to diverge significantly; think about extensibility of the software and use modern tools for version control.

\subsection{Changing Data Layout to for Modern CPU Architectures}
\begin{figure}[ht!]
\includegraphics[trim=0 475 110 0,clip,width=\textwidth]{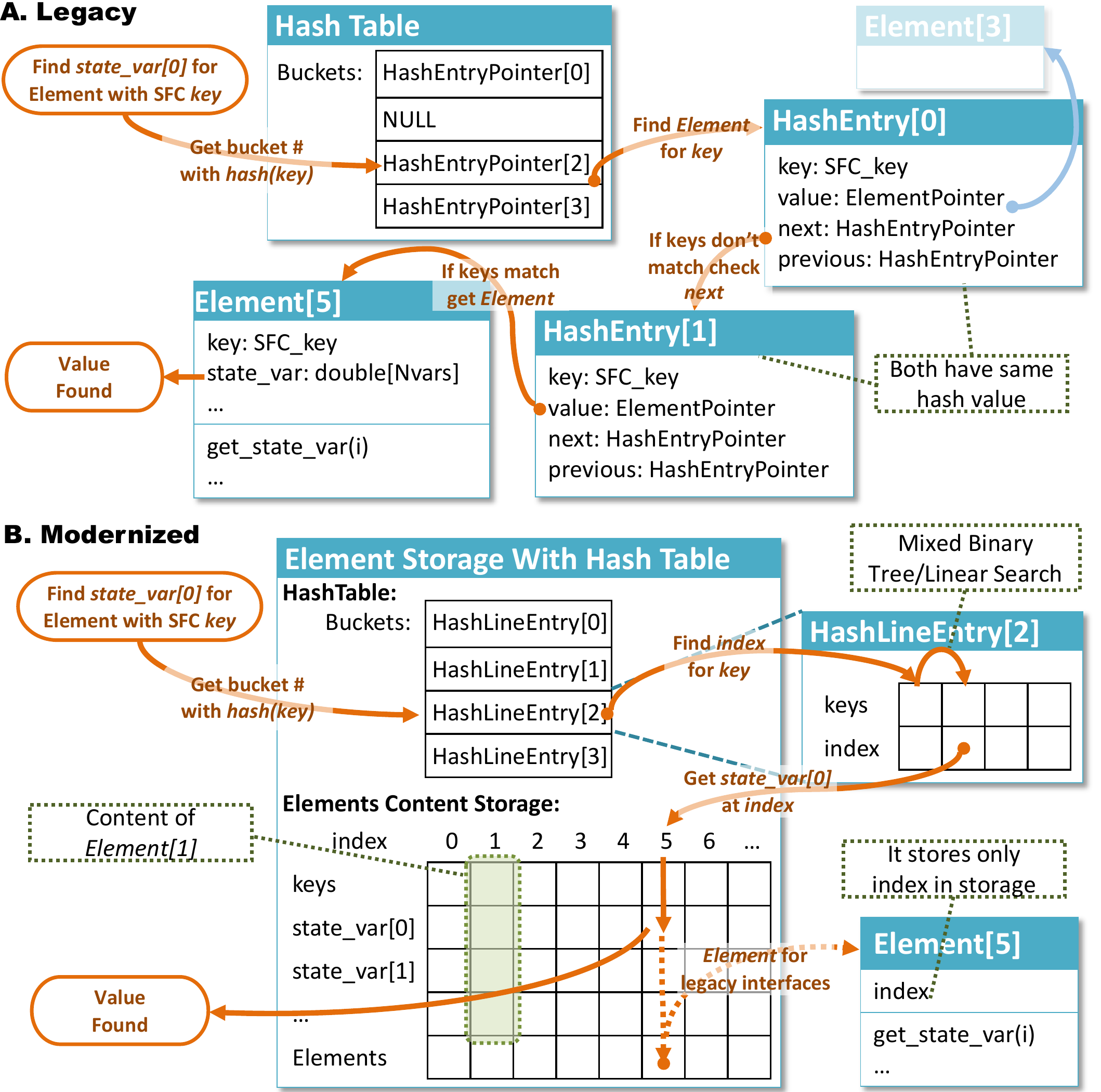}
\caption{Diagram showing search for state\_var value of element with key SFC value. Another common operation is to cycle other all elements: in legacy version it was accomplished by looping through all linked hash entries for each hash bucket, in modernized version it is done by going through array of required data.} \label{figDiagramm}
\end{figure}

Modern CPUs benefit greatly from  vector instructions which require  use of primitive data type arrays (e.g. arrays of integers or doubles) with stride of one for best results. Therefore, for performance reasons, Structure of Arrays (SoA) concept is more preferable than arrays of structure~\cite{Homann2018} which was often used due to better fit of object oriented design. The original Titan2d stores elements and nodes data in allocated on demand objects, the pointers to which are stored in hash tables with a linked list for overflow entries (Fig.~\ref{figDiagramm}.A). This storage method closely resembles a linked list of structures and does not allow efficient use of vector instructions. Therefore,  the first step was to redesign the data layout to a structure of arrays. Although it looks like an enormous amount of work requiring rewriting most of the code, taking advantage of C++ object-oriented design, data encapsulation strategy, modern refactoring tools, and regular expression substitution allows us to do it in a relatively short period of time.

The new data layout utilizes a structure of arrays approach. The hash table class was augmented with elements storage capability where the element's properties are stored as separate arrays and values with same array index belonging to the same element (Fig.~\ref{figDiagramm}.B). To do a seamless transition, the original element class which stores the element properties values are transformed to an element properties accessor class which is a wrapper around new elements storage class and only stores the element index from the elements storage class. The element properties accessor class retains the same API as the original element properties class, allowing the existing functions to use new accessor class instead of old element class without any modifications as long as they do not access elements data directly but through getters and setters methods. Unfortunately, as is the case with many older transitional C to C++ codes even though getters and setters are present, they are not consistently utilized throughout the code. In addition, some of getters and setters operate on pointers. Therefore, prior to a transition to SoA, preparatory work needed to be done to ensure complete data encapsulation and work of getters and setters on proper data reference type.

The overall process of changing the data layout was as follow: 1) reinforce proper data encapsulation throughout the code, 2) create new elements storage class with necessary data manipulation infrastructure and old element class members transfer to new storage class, and 3) update the hot functions to use new storage class directly instead of the element accessor class.

\noindent  \textbf{Data encapsulation reinforcement}. For each member of element and node class we ensure first that the proper data access interface is used. For example, accessor to class member \textit{double state\_vars[NUM\_OF\_STATE\_VARS]} was \textit{double* get\_state\_vars()} which essentially is almost the same as to use data directly and throughout the code it is used for both setting and getting. It was replaced with \textit{double state\_vars(const int idim) const} and \textit{void state\_vars(const int idim, const double value)}. Next the direct class members access was replaced with respective getters and setters methods. Manual refactoring for a large code is a tedious task. Fortunately the refactoring and code analyzing capabilities of modern IDEs significantly simplify interface transformation. NetBeans and Eclipse were used in this work. Unfortunately, these IDEs do not allow automatic conversion from direct members access to access through getters and setters. To aid this task the following strategy was used for each member of element and node classes: first using the IDE’s refactoring tools the class member was renamed to a code-wide unique name to prevent accidental substitution during the next step, then the regular expression substitution was used to replace class member accesses with the proper getter or setter method. The IDE’s regular expression substitution tool allows guided substitution throughout the whole code reducing changes for unintended substitution. The regression tests were run regularly to ensure that refactoring behaved as expected. Overall the described process is a labor intensive task,  however it is tractable with proper tools and strategy.

\noindent \textbf{Elements content storage class}. One of the key features of Titan2D is adaptive mesh refinement. This poses additional challenges on the elements properties storage class, due to the need to insert new elements as well as delete old elements. The linked structures on x86 architecture often have the best performance for this type of operation however their performance suffers during access to a small selection of their content due to a pseudo random access pattern. A structure of arrays approach offer better performance in the latter case but has worse performance for element insertion and deletion. With proper care this performance impact can be drastically reduced.

In the legacy implementation the elements are exclusively referred by their space filling curve (SFC) value. Without getting into detail the SFC function maps 2D coordinates to a single value and with adequate care for numeric accuracy such a mapping is unique. The elements are found through the hash table which uses truncated SFC values as the hash value. When iteration through all elements is needed it is done by iterating though all hash table entries. Thus, for elements insertion and removal one only needs to maintain the hash table and it is possible to use element appending in the elements properties storage class and delayed deletion. Elements are inserted only during the mesh refinement step and removed only during the coarsening step. For sequential storage of elements properties a new container class template, tivector, was created, it is similar to standard C++ std::vector class in that it pre-allocates internal arrays for larger number of entries (capacity) than requested (size) and reallocation happens only if new size exceeds the capacity. When a new element is created its properties are appended to corresponding tivectors and its mapping of SFC value to index is added to the hash table. This way the performance for new element insertion is comparable with a linked list implementation. For the element deletion it needs only be removed from hash table, the actual deletion from storage arrays can happen later in bulk. With the exception of mesh refinement and coarsening steps, the elements properties are stored ordered according to the element’s SFC value. This is done in order to improve memory locality using the locality property of SFC. Points with similar SFC values are located close to each other in 2D space. The SFC values are also used for load balancing during MPI runs. Actual element data deletion and reordering of new elements occurs only after both mesh refinement and coarsening steps are complete. Keeping track of new and removed elements allows us to use it during sorting step and reduce computational cost.

The migration to new element storage class was completed by removing all members from the old element class, adding the index of the element within the storage class and by changing the getters and setters to use the respective values from the storage class. The class member access method was implemented as an inline method allowing most modern compilers to replace the function call with direct class member access. The performance improvement due to changing the layout was 12.9\%.

\noindent \textbf{Updating the hot functions}. The time-consuming functions were identified using Intel VTune profiler. In these functions the access to elements content were replaced with direct access to respective variables from the element content storage class. In most cases due to the unstructured grid the conditions for successful automatic vectorization were not met and thus we did not get significant performance improvement from this particular step. However, in the element storage class there is no strict requirements on the order of stored elements (the ordering by space filling curve values is done mainly for region distribution between MPI processes) and therefore in the future we may be able to reorder the elements to favor vectorized instructions.

\subsection{Efficient Indexing for  Elements/Nodes Addressing}

As mentioned in the previous section, in the legacy implementation the elements are referenced by their SFC values and located through a hash table by their SFC value. The frequent hash table use for lookup presents a significant performance impact even for moderately sized simulations due to high amount of hash values collision (i.e. multiple entries have the same hash value). In the legacy Titan2D these collisions are resolved by a linear search through an overflow linked list of entries.  Titan2D modernization  replaces the overflow linked list entries with usage of standard C++ vectors for hash table buckets. This allows us to implement binary search with linear search at last level. This improvement gives us about 50\% improvement in performance for large problems.

We also realized after the implementation of new elements properties storage class that we can use indexes from the storage class instead of SFC values and thus avoid the use of hash table in many cases. First, we implemented index search prior to getting into the PDE solving routines and later we implemented proper index handling during mesh refinement and coarsening.

\subsection{Introducing OpenMP and Hybrid OpenMP/MPI Parallelization}

Single core performance improvements have declined in recent years, resulting in an increasing number of core per node. It is common nowadays for HPC compute nodes to have half a hundred cores and many performance desktops have eight or more cores. Such high core counts and shared memory space make a multithreaded parallelization approach very attractive. Implementing parallel algorithms with OpenMP is relatively simple, although it can be challenging to achieve best possible performance. Therefore, we decided to go with hybrid OpenMP/MPI parallelization scheme and at first pay more attention to performance of pure OpenMP on a single node while ensuring that MPI parallelization is still working. Previously described modifications are mainly concerning single core performance and by adding OpenMP we following the strategy to maximize single node performance prior moving to multi node runs. OpenMP parallelization was done for the most time-consuming parts of Titan2D. Such parts were identified with Intel Vtune profiler. At our initial approach we largely add pragmas to implement parallel for loops (leaving room for future improvement with more restructuring to take advantage of larger parallel regions, etc.). Due to the large increase in performance of a single process the inefficiencies of MPI implementation become a limiting factor for performance growth in MPI runs to a degree that two MPI processes takes longer time to execute than a serial run. It is not uncommon that improvements in one part of the program reveal inefficiencies on other part as now they become a larger limiting factor.


\begin{figure}[t]
\includegraphics[width=\textwidth]{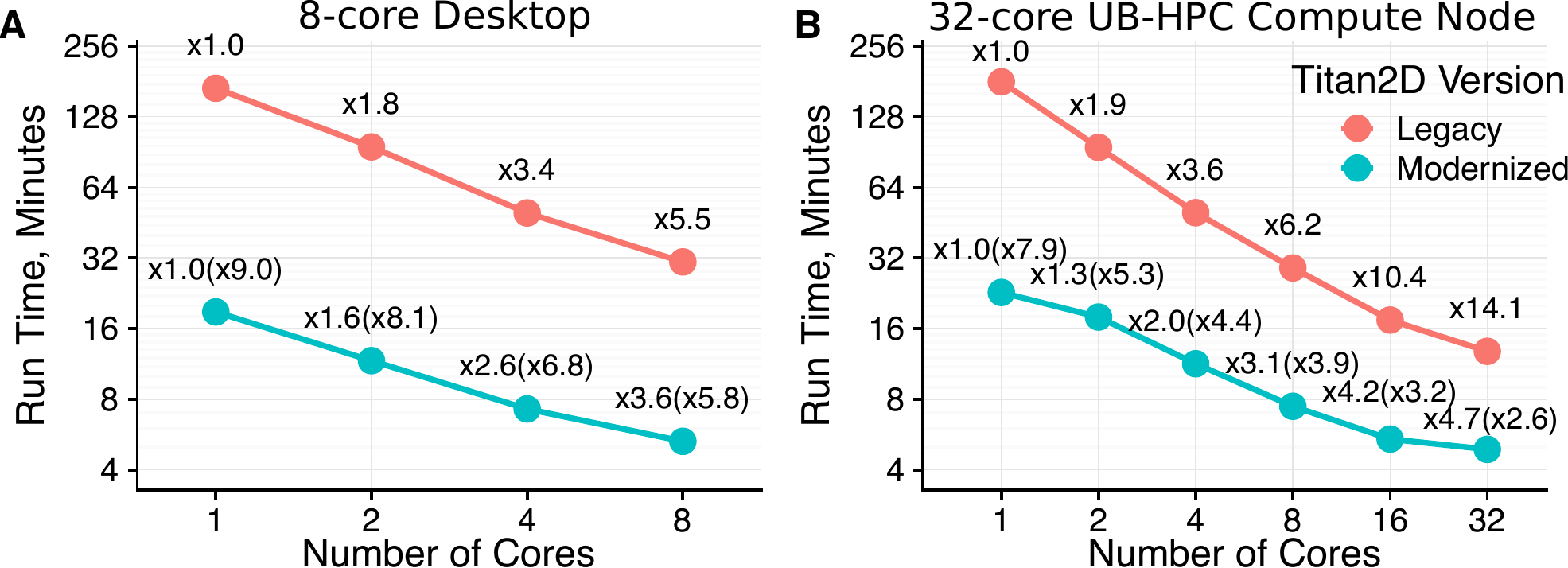}
\caption{Comparison of Titan2D performance as measure by execution time of Colima volcano benchmark between legacy and modernized version executed on \textbf{A} 8-core desktop and \textbf{B} 32-core compute node from UBHPC cluster. The number for each point shows the parallel scaling and the number in brackets shows speed-up from the legacy version executed on same number of cores. Note the $\log_2-\log_2$ scales. For this benchmark, both Titan2D versions are limited to a single node.} \label{figWalltime}
\end{figure}

\section{Performance Improvement Evaluation}

The overall effect of the modifications is shown in Fig.~\ref{figWalltime}. Single core performance is improved by 8 to 9 times, reducing calculation time from hours to tens of minutes. OpenMP parallel run shows a very modest parallel scaling on a single socket: the parallel speed-up is 3.6 on 8 cores, which is smaller than speed-up of 5.5 demonstrated by legacy Titan2D (Fig.~\ref{figWalltime}.A). On a dual socket system the OpenMP parallel speed-up starts to level-off on largest core count (Fig.~\ref{figWalltime}.C). Interestingly, the speed-up on 2 cores is smaller for dual socket system, which is most likely due to lack of NUMA awareness and no CPU affinity reinforcement. 

The not so ideal parallel performance of the modernized version is mainly due to a less than optimal implementation with OpenMP. To date we have mainly concentrated on fork merge and data layout transformation. Still the performance of modernized version on a single core of 8-core desktop is 64\% higher than the performance of the legacy version on all 8 desktop cores and for the dual socket system 4 cores can perform better than the legacy code on the whole node (13\% faster).

\begin{table}[t]
\caption{Comparison of performance metrics obtained with Intel VTune profiler.
Executed on eight cores, no. of elements and iterations were reduced for more tractable profiling.
}\label{tabPerformanceMetrics}
\begin{center}
\begin{tabular}{lrr}
\hline
Metric                              & Legacy      & Modernized \\ \hline
Elapsed time, s                     & 371.7       & 69.2       \\ \hdashline
Instructions retired                & 4.28x$10^{12}$    & 1.41x$10^{12}$   \\ \hdashline
Cycles per instruction rate (CPI)   & 2.84        & 1.50       \\ \hdashline
DRAM bound, \% of clockticks        & 60          & 20         \\ \hdashline
Double precision GFLOPS             & 1.01        & 1.93       \\ \hdashline
MPI busy wait time                  & 0.8s(0.2\%) &            \\ \hdashline
OMP parallel region, \% of walltime &             & 63.0\%    \\ \hline
\end{tabular}
\end{center}
\end{table}

\begin{table}[t]
\caption{Comparison of hotspots obtained with Intel VTune profiler. Number of elements and interation was reduced for more tractable profiling.}\label{tabHotSpots}
\begin{center}
\begin{tabular}{lrlr}
\hline
Legacy               & CPU time, s & Modernized                                  & CPU time, s \\ \hline
HashTable::lookup    & 838.37      & OMP barriers                                & 187.88      \\ \hdashline
MPI data moving      & 366.95      & Integration step                            & 34.08       \\ \hdashline
Integration step     & 149.44      & Slope calculation                           & 30.38       \\ \hdashline
Slope calculation    & 137.25      & Edge states calculation                       & 29.88       \\ \hdashline
Statistics calculation & 133.55      & Z-direction flux calculation                    & 15.49       \\ \hline
Total CPU Time & 2963.3 & Total Time                    & 517.1  \\ \hline
\end{tabular}
\end{center}
\end{table}

The performance metrics, as well as hot spots obtained with Intel VTune profiler, is shown in Table~\ref{tabPerformanceMetrics} and~\ref{tabHotSpots}. As can be seen, the modernized version uses two times less instruction in part due to reducing hash table look-ups and improving its algorithm. The new version also has two times better cycles per instruction rate (CPI), that is it do twice more instruction per cycle and almost twice higher GFLOPS. In both cases, the floating point operation was not packed. The new version also reduces its memory bound. The current version still has high time for OpenMP barriers and on 63\% parallel region. Therefore there is still space to improve OpenMP parallelization further.

\section{Conclusions and Future Plans}

Large amounts of work were done to bring Titan2D up-to-date. Introduction of a Python interface allows easier implementation of different material methods as well as numeric methods under the same source code. Merging several Titan2D forks together allows end users to use the same code for different models and discover which one suits their need. Redesign of the data layout provides better memory access and prepares the code foundation for future improvements. The current, modernized Titan2D, still largely does not use vectorized instructions, which is a non-trivial task for unstructured grids.  We envision that proper ordering of elements within the storage class should allow us to implement a scheme which would benefit from vectorization. Still, modernized code already has significant performance improvement over the legacy version.

%
%
\bibliographystyle{splncs04}
\bibliography{references}

\end{document}